# Evidence of a hidden flux phase in the topological kagome metal CsV$_3$Sb$_5$

Li Yu[1,6,9†*], Chennan Wang[2†], Yuhang Zhang[1,6†], Mathias Sander[3,4], Shunli Ni[1,6], Zouyouwei Lu[1,6], Sheng Ma[1,6], Zhengguo Wang[10], Zhen Zhao[1,6], Hui Chen[1,6,7,9], Kun Jiang[1], Yan Zhang[10], Haitao Yang[1,6,7,9*], Fang Zhou[1,6,9], Xiaoli Dong[1,6,9], Steven L. Johnson[3,4], Michael J. Graf [5*], Jiangping Hu[1,8], Hong-Jun Gao[1,6,7,9], Zhongxian Zhao[1,6,9]

[1]Beijing National Laboratory for Condensed Matter Physics and Institute of Physics, Chinese Academy of Sciences, Beijing 100190, P. R. China
[2]Laboratory for Muon Spin Spectroscopy, Paul Scherrer Institute, 5232 Villigen, Switzerland
[3]Institute for Quantum Electronics, ETH Zürich, 8093 Zürich, Switzerland
[4]SwissFEL, Paul Scherrer Institute, Villigen, Switzerland
[5]Department of Physics, Boston College, Chestnut Hill, MA 02467, USA
[6]University of Chinese Academy of Sciences, Beijing 100049, P. R. China
[7]CAS Center for Excellence in Topological Quantum Computation, University of Chinese Academy of Sciences, Beijing 100190, P. R. China
[8]Kavli Institute of Theoretical Sciences, University of Chinese Academy of Sciences, Beijing, 100190, P. R. China
[9]Songshan Lake Materials Laboratory, Dongguan, Guangdong 523808, P. R. China
[10]International Centre for Quantum Materials, School of Physics, Peking University, Beijing 100871, P. R. China

†These authors contributed equally to this work
*E-mail: li.yu@iphy.ac.cn (L.Y.); htyang@iphy.ac.cn (H.T.Y); michael.graf@bc.edu (M.J.G);

**Phase transitions governed by spontaneous time reversal symmetry breaking (TRSB) have long been sought in many quantum systems, including materials with anomalous Hall effect (AHE)[1,2], cuprate high temperature superconductors[3-8,50], Iridates[9,10] and so on. However, experimentally identifying such a phase transition is extremely challenging because the transition is hidden from many experimental probes. Here, using zero-field muon spin relaxation (ZF-μSR) technique, we observe strong TRSB signals below 70 K in the newly discovered kagome superconductor $CsV_3Sb_5$[11-14]. The TRSB state emerges from the 2 x 2 charge density wave (CDW) phase present below ~ 95 K. By carrying out optical second-harmonic generation (SHG) experiments, we also find that inversion symmetry is maintained in the temperature range of interest. Combining all the experimental results and symmetry constraints, we conclude that the interlayer coupled chiral flux phase (CFP)[15] is the most promising candidate for the TRSB state among all theoretical proposals of orbital current orders[15,33, 40, 42-44]. Thus, this prototypical kagome metal $CsV_3Sb_5$ can be a platform to establish a TRSB current-ordered state and explore its relationship with CDW, giant AHE, and superconductivity.**

The two-dimensional kagome lattice has attracted much interest in the condensed matter community because it can host a variety of exotic electronic physics, such as quantum spin liquids, flat-bands, various topological states and so on[16-23]. Recently, a new family of kagome superconductors $AV_3Sb_5$ (A = K, Rb, Cs) with nontrivial $Z_2$ band topology has been discovered [11-14]. Its crystal structure is formed by a succession of alternative stacking of an alkali-metal layer and a $V_3Sb_5$ layer, where the vanadium atoms form a quasi-2D kagome lattice (Fig. 1a, b, c).

The materials display very intriguing physical properties. A charge-density-wave (CDW) with a 2 × 2 order appears at an onset temperature $T_{CDW}$ ranging from 80 K to 110 K[11, 12, 14, 26]. A giant anomalous Hall effect (AHE) is observed along with a magnetic field induced CDW chirality at lower temperatures [30,35][13, 28], and the weak onset of AHE is concomitant with the CDW order. These peculiar phenomena suggest that the time-reversal symmetry may also be broken. However, both neutron scattering[11] and µSR experiments[29] have demonstrated that there is no measurable local spin moments or magnetic correlations in $AV_3Sb_5$[11, 13, 14, 29, 37]. Namely, the potential TRSB cannot originate from magnetic ordering. Several models of orbital currents [15, 33, 40, 42-44] with different closed-loop patterns have been theoretically proposed to explain the TRSB. While the TRSB states with orbital currents have been proposed in many topological or correlated electron systems[1-10], to our knowledge, they are extremely difficult to confirm directly by most experimental measurements.

The µSR technique is a uniquely sensitive probe of TRSB because it can detect the strength and spatial distribution of the intrinsic local magnetic field at the muon stopping sites inside a bulk material. A previous µSR experiment has shown a weak and static magnetism developing in polycrystalline $KV_3Sb_5$ sample[29], where a similar charge ordering appears below 80 K[14]. However, the nature of this weak magnetic signal was left unresolved. Especially, it is intriguing to ask whether the signal reflects a particular orbital current state.

Here, we perform more detailed µSR investigations on single crystals of $CsV_3Sb_5$. As summarized in Fig. 2**a**, we have identified a spontaneous breaking of the time-reversal symmetry at a characteristic temperature $T^* \sim 70$ K, which is well below the charge ordering temperature $T_{CDW} = 95$ K. Information on the distribution of local magnetic field ($\boldsymbol{B}_\mu$) at the muon stopping sites can be extracted from time evolution of the muon spin polarization $\boldsymbol{p}_\mu(t)$ transverse (or longitudinal) to the local field (see supplemental material S2). The temperature evolution of the local magnetic field distribution can be accessed by monitoring the $T$-dependent depolarization rate ($\sigma_\mu(T)$). Subtle changes of the local field distribution even by a very small magnitude of 0.01 Oe can be detected. Experimentally, µSR measures $\boldsymbol{p}_\mu(t)$ projected in the **c**- or **ab**-directions, as shown in Fig.2**b**, enabling the local field anisotropy to be identified at the muon sites, thus revealing the structure of the local field ($B_\mu$) distribution. As a simplified description of how such a connection is established, the finite depolarization rate with $p_\mu$//**c** measures predominately the transverse field distribution components in the **ab**-plane ($B_{//}$), while a finite depolarization rate along $p_\mu$//**ab** is associated with the field components from the **ac**- or **bc**-plane ($B_\perp$).

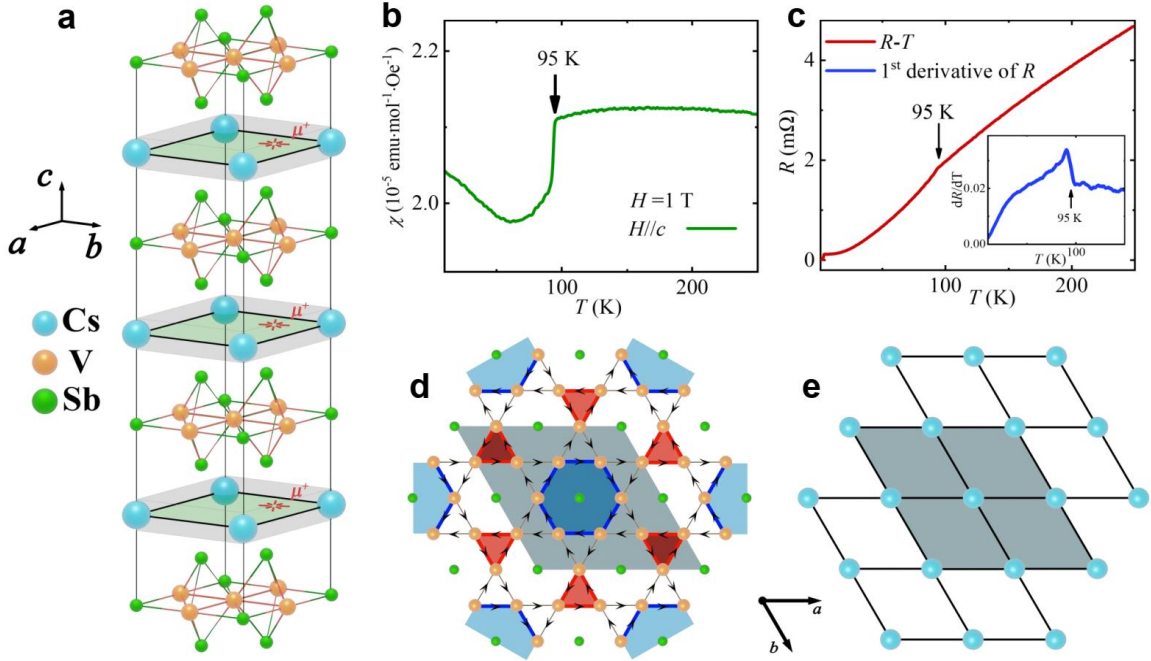

**Fig. 1: Crystal structure of CsV$_3$Sb$_5$, muon stopping sites, magnetization and transport characterizations, and sketch of the kagome and Cs plane in the chiral flux phase. a,** schematic drawing of the lattice structure of CsV$_3$Sb$_5$. The approximate muon stopping sites are in the close vicinity to the Cs hexagonal plane (gray area) within each unit cell. **b**, Magnetic characterization under $H$ = 1 T along **c**-axis of the CsV$_3$Sb$_5$ single crystal sample used in our µSR experiments. Abrupt reductions of spin susceptibility due to the formation of CDW order around 95 K are clearly observed. **c**, The in-plane resistance (red) and corresponding 1$^{st}$ derivative (insert blue) are presented with kink (peak) like anomaly onset around 95 K. **d**, schematic drawing of kagome lattice V$_3$Sb$_1$ with order of orbital current according to the chiral flux phase model[15]. **e**, a clear representation of the corresponding alkali metal planes. The gray areas stand for the extended 2 x 2 unit cell of the CDW order.

The temperature evolutions of the normalized muon depolarization rates [$\sigma_\mu(T)$- $\sigma_\mu(150\ K)$]/ $\sigma_\mu(150\ K)$ obtained along the **ab**-plane and **c**-axis are shown in Fig. 2**a**. With decreasing temperature, the normalized rates for both polarizations stay nearly constant crossing $T_{CDW}$ = 95 K, however, a monotonic increase of the signal absolute amplitude is observed at temperature below $T' \sim$ 70 K. This indicates that a distinct transition emerges well below

$T_{CDW}$, which has been hidden from most previous experiments. Below $T'$, the two muon depolarization rates for $p_\mu$//**ab** and $p_\mu$//**c** geometries evolve in the opposite manners, with the one for $p_\mu$//**c** enhanced significantly, and the other for $p_\mu$//**ab** slightly decreased with cooling. Such a significant discrepancy is a clear evidence for the local anisotropy of internal magnetic field. To be more specific, the enhancement of the muon depolarization rate for $p_\mu$//**c** is due to the appearance of local transverse fields $B_\mu$//**ab** in the crystallographic **ab**-plane, whereas the same local field naturally contains longitudinal component with respect to the muon spin polarization for $p_\mu$//**ab**, causing the reduction of the muon depolarization rate. This is expected since the longitudinal fields decouple or dwarf the local magnetic environment from the muon spin[45]. Upon cooling to $T'' \sim 30$ K, both depolarization rates start to show concurrent increases, which implies a switching into a different local magnetic field distribution. At temperatures below ~15 K, an apparent saturation of the depolarization rate occurs. A slightly enhanced muon relaxation rate below the superconducting transition temperature further implies that the internal field is static and becomes inherently inhomogeneous due to Meissner screening at zero-field environment. Therefore, we identify a spontaneous breaking of the time-reversal symmetry at the characteristic temperature $T' \sim 70$ K using µSR. We emphasize that the onset of this hidden TRSB phase does not coincide with the CDW transition at $T_{CDW} = 95$ K, as indicated in Fig.2**a**.

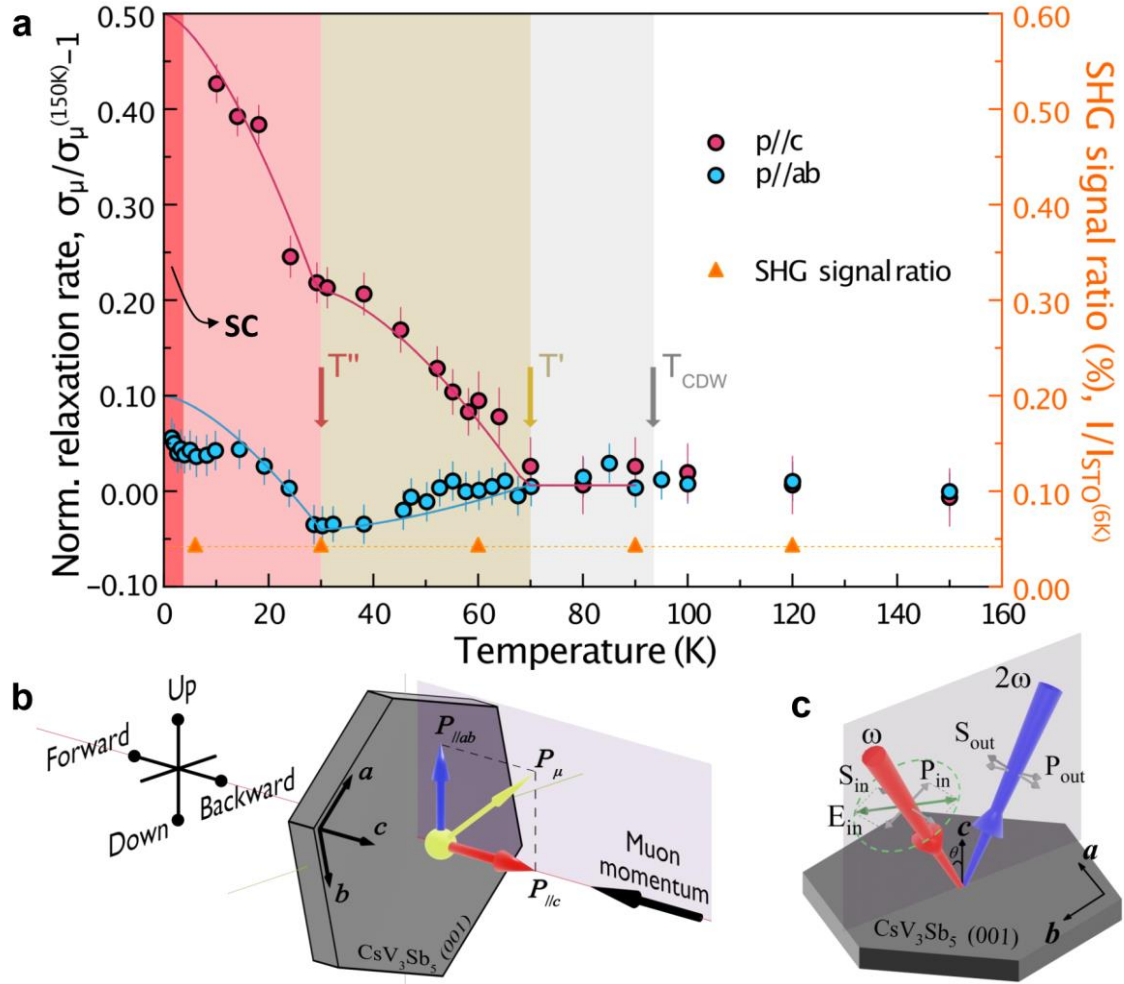

**Fig. 2: Identification of the time-reversal symmetry breaking and checking the lattice inversion symmetry in CsV₃Sb₅. a**, diagram illustrates distinct time-reversal symmetry breaking phases, determined by temperature dependence of (left black vertical axis) the anisotropic muon spin depolarization rate along Forward-Backward (F-B) and Up-Down (U-D) projections from the initial muon polarization ($p_\mu$ for initial muon spin polarization). Both rates are normalized to the mean value around 150 K. A clear onset appears at $T' \sim 70$ K which is distinct from the charge ordering temperature $T_{CDW} \sim 95$ K, as marked in the plot. Curves are guides to the eye represent by a quadratic function in order to highlight the transition temperature. Below $T'' \sim 30$ K, both depolarization rates possess simultaneous enhancement with decreasing temperature, suggesting the change of the local magnetic structure. (right orange vertical axis) The second harmonic generation (SHG) signal, whose absolute intensity is normalized with the maximum signal of the reference SrTiO₃ sample at 6 K under the same incidence beam fluence. The SHG signal generated by the CsV₃Sb₅ sample is not distinguishable against the weak signal from the measurement background. **b**, the schematic drawing of the μSR measurement geometry used in this work. The muon beam is normal to the sample **ab**-plane. The initial muon spin $p_\mu$ is rotated from the incident muon moment direction with projected muon spin components along F-B (red filled circle,

$p_\mu$//**c**) and U-D (light blue filled circle, $p_\mu$//**ab**) directions. **c,** geometry of the SHG experiment, where the incidence angle θ is fixed at 45 degrees.

The TRSB observed in CsV$_3$Sb$_5$ is quite prominent. As there is no measurable short- or long-range magnetic order in CsV$_3$Sb$_5$, the TRSB cannot originate from any conventional magnetism. Therefore, it is natural to ask whether our μSR results are consistent with the proposed loop current states[15,33,40,42-44]. In particular, we are interested in whether our result can be helpful to determine the specific state of orbital current. According to a recent theoretical classification, there are 18 classes of possible orbital current states with 2 × 2 in-plane order[53]. In order to pin down the genuine one, we use symmetry principles to check the remaining symmetry of this hidden phase by applying the second-harmonic generation (SHG) technique to CsV$_3$Sb$_5$. SHG measures the second-order nonlinear optical response $\boldsymbol{P} = \varepsilon_0 \chi^{(2)} \boldsymbol{EE}$, where $\boldsymbol{P}$ is the electric polarization induced by the incident light with electric field $\boldsymbol{E}$ and $\varepsilon_0$ is the vacuum permittivity, as illustrated in Fig. 2**c**. Since $\boldsymbol{P}$ and $\boldsymbol{E}$ are odd under inversion symmetry, the rank-three nonlinear optical susceptibility tensor $\chi^{(2)}$ is only finite when parity is broken. Our SHG experimental results are summarized in Fig. 2**a**. From 120 K down to 6 K, we only observed negligibly small amplitude of signals against the measurement background. Thus, no signature of the lattice inversion symmetry breaking[48,49] is detected at all measured temperatures.

The above result provides a strong symmetry constraint to possible loop-current states. CsV$_3$Sb$_5$ crystalizes in $D^1_{6h}$ (P6/mmm) space group[31]. Below the CDW transition, the "Tri-Hexagonal" or "Star of David" structure deformations are supposed to occur. These two structures still keep the $D_{6h}$ while enlarging the unit cell to 2 × 2 (Fig.1**b** and 1**c**) in the **ab**-plane. At lower temperatures above the superconducting (SC) transition, the $C_6$ rotational symmetry was also found to be reduced to $C_2$ by STM and transport measurements[32]. Combined with the maintained inversion symmetry, we can conclude that the point group symmetry of this hidden TRSB phase is at least $C_{2h}$. We note that this analysis is consistent with the appearance of all commensurate interlayer displacement[25] induced 2 × 2 × 2 CDW pattern. Combining the experimental results and the theoretical classification, only the $D_{6h}$ flux phases are possible configurations for the hidden TRSB phase. Among the 18 flux classes[53], there are only 4 possible classes to satisfy this constraint. Furthermore, the chiral flux phase (CFP)[15] has been theoretically shown to have the lowest energy among these 4 classes.

We find that the presence of the hidden CFP state can be further elaborated from our experimental data. We notice that there is an aforementioned significant anisotropy in the muon depolarization rates along both $p_\mu$//**ab** and $p_\mu$//**c** directions. This anisotropy can be produced by constructing the microscopic configuration based on the symmetry analysis of the local fields at the muon stopping sites[24]. According to calculations from previous μSR work[29], the predominating muon sites are expected to be located in close vicinity to the alkali metal planes, which maximizes the distance with respect to the V and Sb atoms and centers itself in between two neighboring kagome layers. (Fig.1**a.** The gray areas close to each Cs-planes.) The local magnetic field probed by muon at the stopping sites of such symmetry is the vector sum of the stray fields from the loop fluxes of all kagome layers above and below the Cs-plane. Notably, the local field is also very sensitive to any imbalance between magnetic structures of kagome planes on either side. Considering the big lattice spacing along **c**-axis (> 9 Å) and thus strong decay of the stray field from the loop fluxes, we only need to consider a few kagomé layers close to each Cs-plane. Here, we will use the smallest building block, "Bilayer-Kagome configuration" (BKC), to demonstrate the interlayer magnetic coupling for modeling the local field distribution close to the Cs plane. as illustrated in Fig. 3.

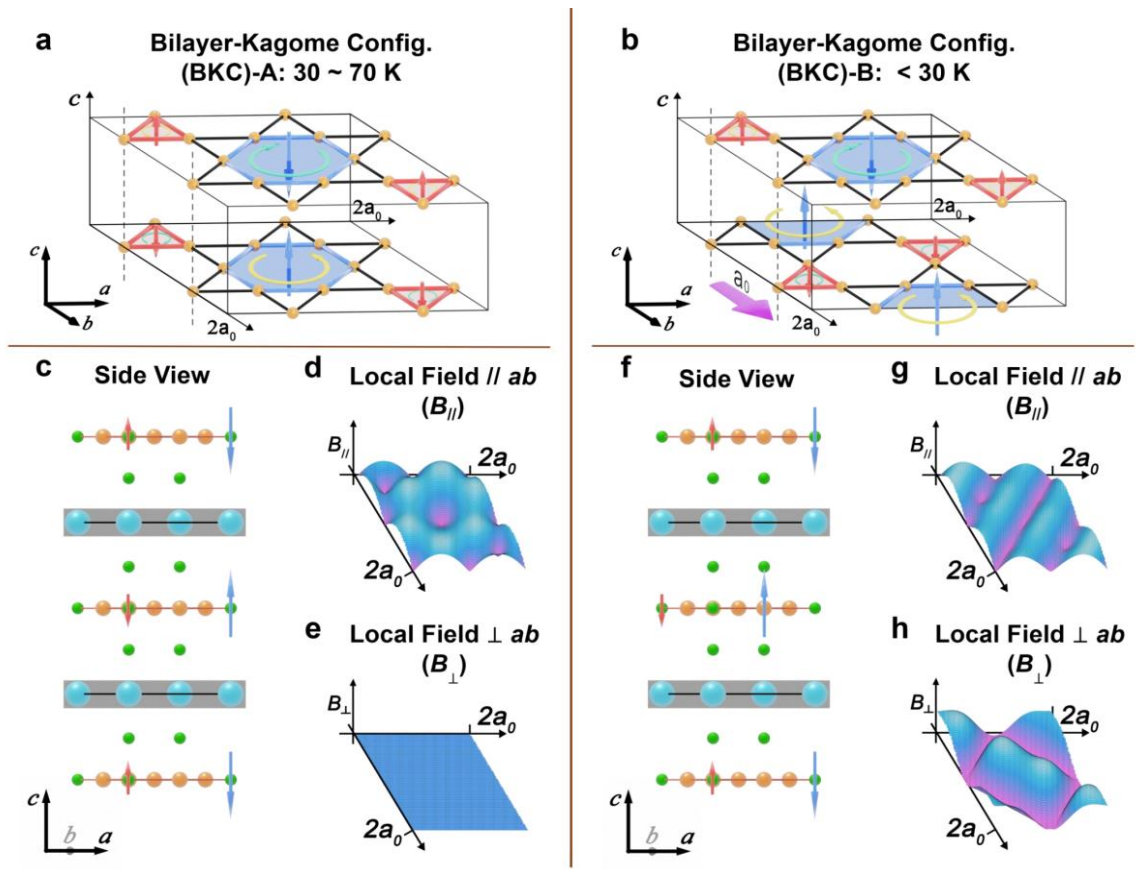

**Fig. 3: Simulation of local field distribution along the Cs-plane within the unite cell of hidden CFP order.** In **a,** and **b,** are schematic drawings for demonstrating the interlayer magnetic configuration of nearest neighboring kagome planes for Bilayer-Kagome configuration (BKC) - A: for 30 ~ 70 K and BKC- B: for < 30 K, respectively. The length of plotted arrow is in proportional to the size of the effective orbital current loop flux strength. Corresponding sideview from ac-plane of the magnetic field configurations on the kagomé lattice are shown in **c**, and in **f**. Here the anti-phase magnetic structure on the nearest neighboring kagomé planes in **c** has one lattice constant $a_0$ shift along one of the crystallographic axis with respect to the neighboring planes in **f**, indicated by the purple arrow. Both magnetic configurations fulfill with a 2 x 2 x 2 magnetic unit cell for the hidden CFP state. On the right side accordingly, **d-e** and **g-h**, are the corresponding calculated in-plane ($B_{//}$) and out-of-plane ($B_\perp$) local field distribution along the Cs-plane. The preserved inversion symmetry of the crystal lattice is maintained at all temperatures according to the negligibly small SHG signal (see text). (Here for the simulation of local field, we only consider two kagomé layers on either side of every Cs-plane, which is enough to correctly calculate the local field distribution on the Cs-plane)

To understand such anisotropy, we simulate the field distribution sensed by muons in the CFP state[15]. As shown in Fig. 1**b**, there are two distinct orbital currents in the CFP state, looping along the triangle (red-shaded area) and hexagonal (blue-shaded area) sub-lattices, respectively, in opposite flowing directions. The loop fluxes, serving as effective local moments, are anti-parallel to each other and double the in-plane unit cell as marked by gray areas in Fig. 1**b** and 1**c**. Along the c-axis, the CFP state can have two different configurations, which can result in two different BKCs as shown in Fig.3**a** and 3**b**. In the first BKC, BKC-A (Fig. 3**a**, and 3**c**), it possesses a magnetic structure on two neighboring kagome layers with an out-of-phase interlayer magnetic coupling. In the second BKC, BKC-B (Fig. 3**b**, and 3**h**), the anti-phase magnetic structure between the layers is shifted by one lattice constant $a_0$. The corresponding ac-plane sideview of the magnetic unit cell for both BKC configurations are presented in Fig. 3**c** and 3**f**, respectively. The in-plane ($B_{//}$) and out-of-plane ($B_\perp$) components of the local field distributions on the Cs-plane within the CFP unit cell (shaded areas in Fig. 3**c** and 3**f**) are shown on the right side.

We find that the CFP model simulation of the BKC-A and BKC-B configurations are highly consistent with the local field distribution at $T = 30 \sim 70$ K and $T < 30$ K, respectively. For BKC-A, the simulation reveals an anisotropic local field structure in the Cs-plane with a vanishing $B_\perp$ component and a finite $B_{//}$ component. This is compatible with the in-plane local field revealed by anisotropic muon depolarization rates in both $p_\mu//c$ and $p_\mu//ab$ geometries as shown in Fig. 2**a**. Below 30 K, instead, the BKC-B local field simulation is consistent with our experimental results. In this case, both muon depolarization rates in $p_\mu//ab$ and $p_\mu//c$ are increased simultaneously as shown in Fig. 2**a**, due to the appearance of sufficient local field in both in-plane and out-of-plane directions. It is noteworthy that such clear consistency between the calculated local field distribution and our μSR observation is owing to the muon site symmetry and importantly, the anti-phase interlayer magnetic configuration. (In the supplemental materials S3, we show that the in-phase interlayer coupling contradicts our μSR observations.) Especially, this anti-phase interlayer magnetic coupling between the two neighboring kagome planes is likely to be essential in the formation of the hidden TRSB order below the characteristic temperature $T^* \sim 70$ K, which is beyond the original CFP model[15]. Nevertheless, the spontaneous internal magnetic structure for the hidden CFP state below 70 K provides us an alternative explanation for the giant AHE in CsV$_3$Sb$_5$ at low $T$[15,28]. In fact, the AHE signal is weak at CDW onset $T_{CDW}$ ~ 95 K and significantly enhanced at lower temperatures.

Our μSR results clearly reveal the two transitions at $T'$ = 70 K and $T''$ = 30 K that are "hidden" from most experimental means. To the best of our knowledge, only the previous transient pump-probe spectroscopy measurements have suggested these two temperature scales: the collective excitation modes at 3.1 THz appearing below 70 K and 30 K, while the CDW transition is hallmarked by a coherent phonon mode at 1.3 THz appearing below ~ 95 K[36, 39]. Other possible experimental hints also exist. For instance, a uniaxial order was observed below about 60 K by scanning tunneling microscopy (STM)[32; 34; 38]. In addition, our ARPES measurements[41] reveal a higher energy scale of 60 meV popping out below 60 ~ 70 K in the EDC spectra around the M-point close to the Brillouin zone boundary, while the CDW order shows a much smaller energy gap of 20 meV appearing at $T_{CDW}$ ~ 95 K. And an extra steeper loss of spectral weight below 30 K in the EDC spectra near Fermi level ($E_F$) has also been resolved. (supplementary materials S4) All these results are consistent with our μSR measurements and reveal unusual electronic correlations, apart from that of the CDW order, that may underlie the exotic hidden order developing below $T'$ ~ 70 K.

In summary, by using zero-field μSR, combined with optical second harmonic generation, we have experimentally identified a hidden orbital current order in the kagome metal CsV$_3$Sb$_5$. Our experimental results suggest that the interlayer magnetically coupled CFP states in kagome layers seem to be the most promising TRSB state. The hidden TRSB phase appears well below the CDW transition, suggesting a distinct origin from the CDW order. An extra magnetic transition below 30 K in the TRSB state has been observed. Furthermore, our simulated local field distributions based on the Bilayer-Kagome configurations of the CFP model are highly consistent with our μSR results. This clear identification of the spontaneous hidden TRSB state in the normal state expands our views on the rich physics in this new quasi-2D kagome metal system. It reveals the kagome metal CsV$_3$Sb$_5$ to be a fascinating class of material to explore the novel physics related to the orbital current states, though the issue of how the hidden flux phase, is connected with the CDW order, and superconductivity warrants further investigations.

**Methods**
**Single-crystal growth and preparation:**
High quality single crystals of CsV$_3$Sb$_5$ were synthesized using the self-flux method[32], and the superconducting transition temperature was characterized as $T_c$ ~3.5 K[27] by resistivity and magnetic susceptibility measurements. The CDW transition temperature ($T^*$) was determined to be 95 K from the clearly visible kink in the resistivity measurements and abrupt reduction in magnetization measurements (see Fig. 1 and Supplementary material S1).

**µSR measurement:**

The uSR experiments on a mosaic of single crystals with **c**-axis coaligned were performed on the GPS spectrometer at the Swiss Muon Source at the Paul Scherrer Institute.[54] The surface muons produced at the target with > 95% spin polarization are implanted in the sample. Muons thermalize rapidly and stop preferential sites of the crystal lattice with the preserved initial muon polarization. The detection scheme of the spectrometer was set to veto-mode to ensure the positron events generated by the muons missing the sample were properly accounted. The measurements for p//**ab** and p//**c** were performed in the muon spin rotated and longitudinal mode, respectively, to reach the optimum measurement conditions. In the spin rotated mode, the initial muon spin rotation angle was about 60 degrees with respect to the muon momentum direction, whereas in the longitudinal mode, the initial muon spin was nearly in parallel with the c-axis of the crystal. During zero field (ZF) measurements, the environmental magnetic field was actively compensated to be below 0.02 Oe at the sample position.

**µSR data fitting procedure:**

The measured µSR spectra were analyzed with the musrfit package[55] and the data is shown in Supplementary S2. To analyze the muon spectra accurately, we adopted the single histogram fit method for directly fitting our muon depolarization model to the positron histograms, allowing each detector efficiency to be treated independently. The positron histogram on the j-th detector $N_j(t)$ is given by: $N_j(t) = N_{0,j} e^{-t/\tau_\mu} [1 + A_{0,j} P_j(t)] + N_{bkg,j}$. Here, $N_{0,j}$ is the measured muon decay events at delay time t=0. The exponential term accounts for the radioactive decay of muon ensemble. $A_{0,j}$ is the instrumental asymmetry of detector pairs. $P_j$ is the projection of muon polarization on the respected detector pairs. $N_{bkg,j}$ accounts for the background events of each detector. A phenomenological Gaussian Kubo-Toyabe depolarization function is used for describing a static and anisotropic field distribution, as given by: $P_j(t) = f_s [\frac{1}{3} + \frac{2}{3}(1 - \sigma_j^2 t^2) e^{-\sigma_j^2 t^2/2}] + (1 - f_s) e^{-\lambda t}$, where $f_s$ describes the fraction of the sample signal, $\sigma_j$ is the muon spin depolarization rate, and $\lambda$ is the relaxation rate of the background contribution.

**SHG measurement:**

The SHG experiments were performed in reflection geometry using a custom-built setup. The incident laser beam was delivered by a Ti:sapphire regenerative amplifier (pulse width

120 fs, repetition rate 1 kHz, center wavelength 800 nm). The scattering plane (oblique incidence angle of 45 deg.) and polarization configuration is illustrated in Fig. 1(e). The polarization of the incidence beam was controlled by rotating a zero-order achromatic half-wave plate. The beam was focused to a 300 μm diameter spot on a freshly cleaved optically flat surface of the sample. A minimum fluence around 10 mJ cm$^{-2}$ was reached at the sample. The CsV$_3$Sb$_5$ and reference SrTiO$_3$ crystals were mounted side-by-side in a closed-cycle cryostat with a pair of quartz optical windows. The reflected beam then passed through a short-pass filter, where the fundamental beam was filtered out and the nonlinear optical signal was detected by a silicon avalanche photodetector.

**Data availability:**
The authors declare that the data supporting the findings of this study are available within the article and its Supplementary Information. All raw data are available from the corresponding author upon request.

**Notes added:**
While preparing this manuscript, another μSR work report on the study of single crystal KV$_3$Sb$_5$ [arXiv2106.13443 (2021), https://arxiv.org/abs/2106.13443] has also revealed a similar TRSB onset temperature at about 70 K, where the CDW onset temperature $T_{CDW}$ = 80 K for KV$_3$Sb$_5$.

**Acknowledgements**

This work is supported by the National Science Foundation of China (Grant Nos. 11834016, 11888101, 12061131005, 51771224, 61888102, 52022105), the National Key Research and Development Projects of China (Grant Nos. 2017YFA0303003, 2018YFA0305800, 2019YFA0308500), the Key Research Program and Strategic Priority Research Program of Frontier Sciences of the Chinese Academy of Sciences (Grant Nos. XDB33010200, XDB25000000, XDB33030100, and QYZDY-SSW-SLH001). This work is based on experiments performed at the Swiss Muon Source SμS at the Paul Scherrer Institute, Villigen, Switzerland.

**Author contributions**

L.Y., H.Y., M.J.G., C.W. conceived the project. H.Y. and Z.Z. synthesized the single crystals. S.N., S.M., Y.H.Z., and Z.L. characterized the samples instructed by X.D., Y.H.Z. and L.Y. conducted the BKC model simulation. L.Y. and C.W. performed the μSR measurements, Y.H.Z., L.Y., C.W. analyzed the μSR data with help from M.J.G., C.W. and M.S. carried out the SHG measurements and analyzed the SHG data with the help of S.J., Z.W., S.M., and Y.Z. performed the ARPES measurements and analysis. L.Y., C.W., J.H., K.J., F.Z. and Y.H.Z. wrote the paper with helpful input from all authors. X.D., S.J., J.H., H.-J.G., and Z.X.Z. supervised the project.